\documentclass[]{emulateapj}%\usepackage{natbib}
\usepackage{ifthen}
\newboolean{astroph}
\setboolean{astroph}{true}

\newcommand{\citeGladman} { (Gladman et al. 2009, in preparation) }

\newcommand{\Msun}{\ensuremath{\mathrm{M}_\odot}}

\newcommand{\pe}{\ensuremath{\mathrm{e}^-}}
\newcommand*{\dif}{\ensuremath{\mathrm{d}}}

\newcommand{\invs}{\ensuremath{\mathrm{s}^{-1}}}

\newcommand{\kmps}{\ensuremath{\mathrm{km\ s}^{-1}}}
\newcommand{\Fsups}{\ensuremath{\mathrm{Fsu\ s}^{-1}}}

\newcommand{\pdegsqrd}{\ensuremath{\mathrm{deg}^{-2}}}
\newcommand{\invFsu}{\ensuremath{\mathrm{Fsu}^{-1}}}
\newcommand{\Fsu}{\ensuremath{\mathrm{Fsu}}}
\newcommand{\fny}{\ensuremath{f_{\mathrm{Ny}}}}

\newcommand{\rKBO}{\ensuremath{r_{\mathrm{KBO}}}}

\newcommand{\dKBO}{\ensuremath{d_{\mathrm{KBO}}}}

\newcommand{\mm}{\ensuremath{\mathrm{m}}}
\newcommand{\mAU}{\ensuremath{\mathrm{AU}}}

\newcommand{\rearth}{\ensuremath{r_\oplus}}
\newcommand{\rocc}{\ensuremath{r_\mathrm{o}}}

\newcommand{\qS}{\ensuremath{q_{\mathrm{S}}}}
\newcommand{\qL}{\ensuremath{q_{\mathrm{L}}}}

\newcommand{\figFTnoise}{
  \begin{figure}[htbp]
    \centering
    \plotone{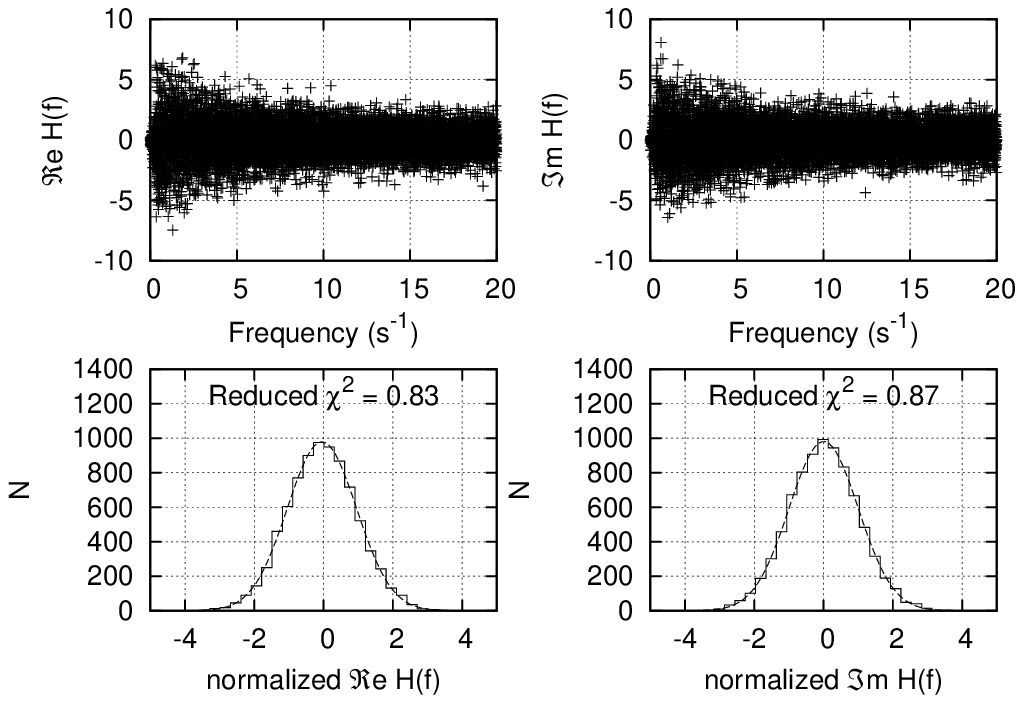}
    \caption
    {The noise distributions of the real and imaginary parts of
      $F(\omega)$.  Upper panels show the real (left) and imaginary
      (right) frequency components.  The frequency components were
      normalized by the standard deviation of points within $\pm0.5
      \invs$, and are shown as histograms in the lower left (real) and
      right (imaginary) panels with best-fit Gaussians.  The frequency
      components are consistent with Gaussian variates based on
      reduced $\chi^2$ values.}
    \label{fig:FTdistrib}
  \end{figure}
}

\newcommand{\figPowerSpec}{
\begin{figure}[htbp]
  \centering
  \plotone{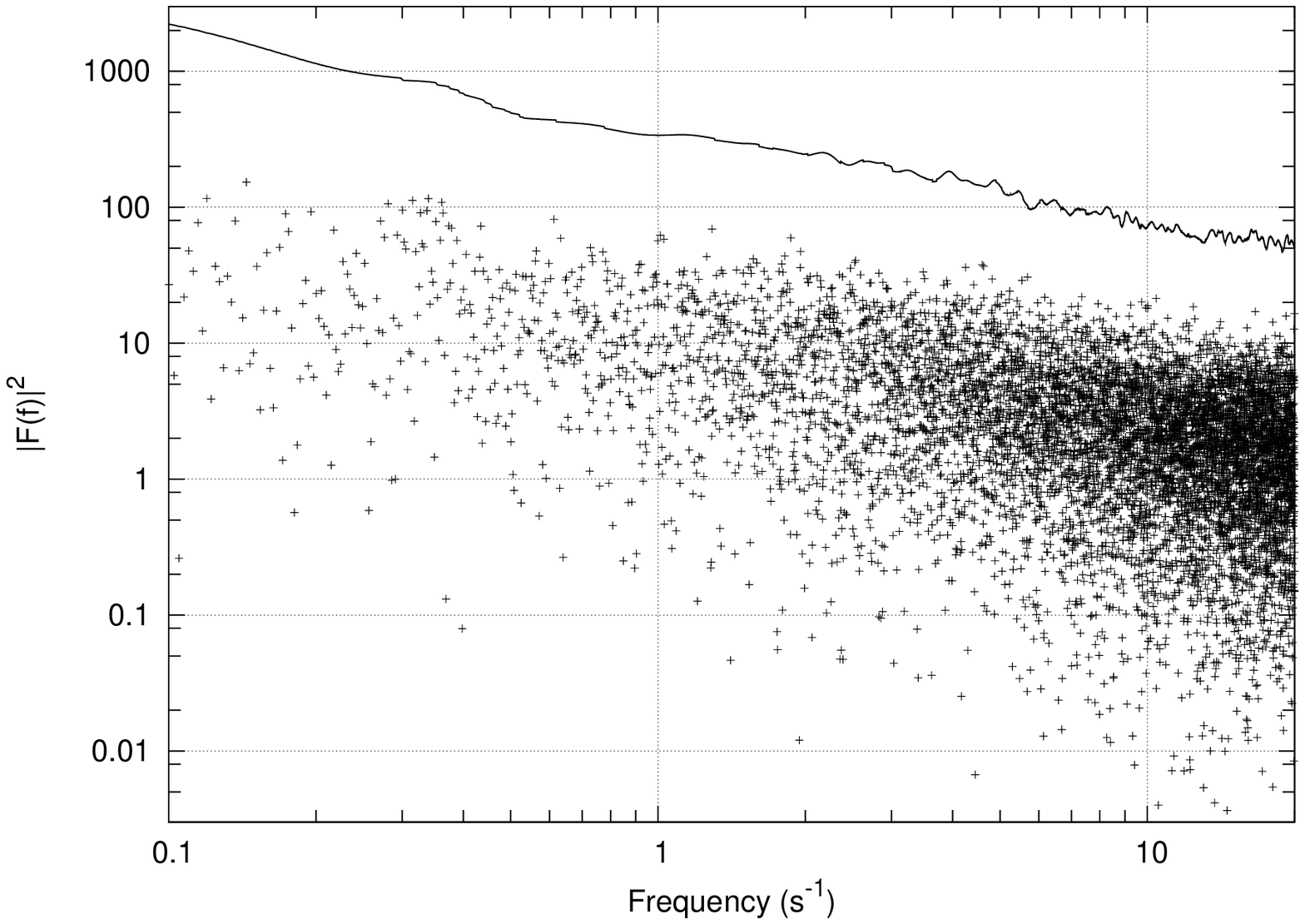}
  \caption
  {The power spectrum for a 40 \invs photometric time series showing a
    power law relationship of $|F| \propto f^\beta$, with
    $\beta\approx1$.  The trend of the average power (0.2 \invs bins)
    is shown offset above for clarity.}
  \label{fig:powerspectrum}
\end{figure}
}

\newcommand{\figFTfake}{
  \begin{figure}[htbp]
    \centering
    \plotone{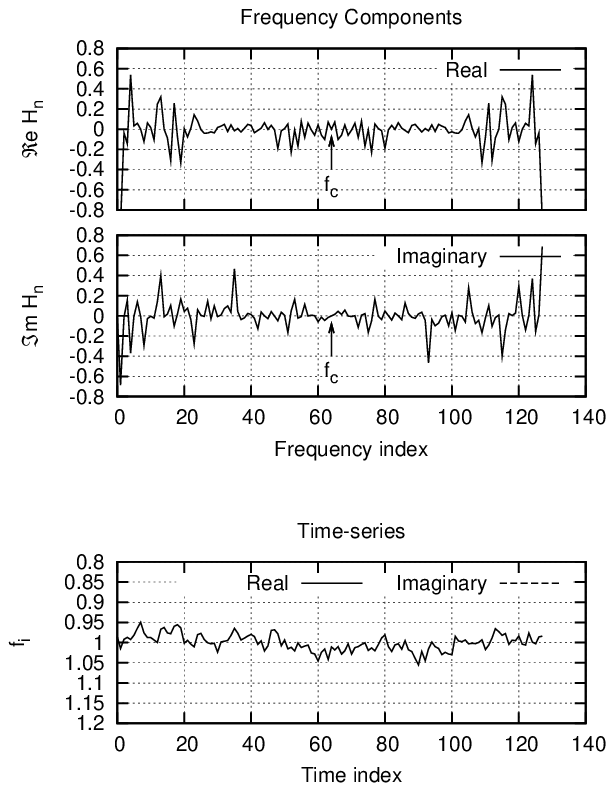}
    \caption
    {Artificially-generated real (top) and imaginary (middle)
      frequency components, and the corresponding time series
      (bottom).  A complex time series is produced, but the imaginary
      components all have values of zero due to conjugate symmetry.}
    \label{fig:FTfake}
  \end{figure}
}

\newcommand{\figOccrates}{
  \begin{figure}[htbp]
    \centering
    \plotone{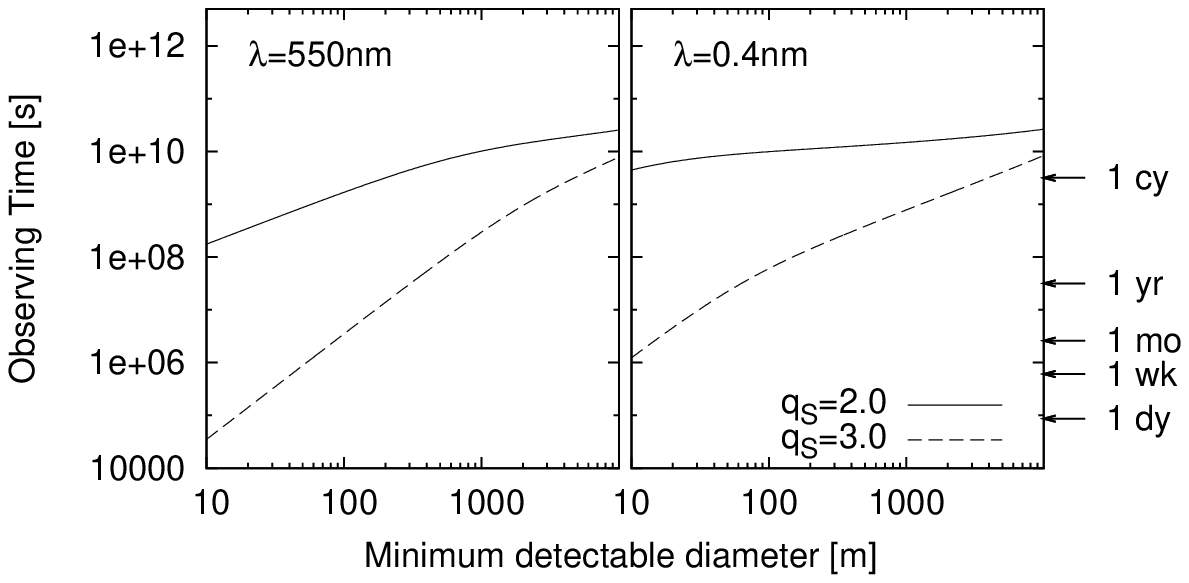}
    \caption
    {Average time between occultations for KBO occultation searches
      performed in the optical (left) and X-ray (right).  Different
      curves represent different possible slopes for the small object
      size distribution, \qS.  The waiting times required for 95\%
      confidence are $-\ln|0.05| \approx 3 \times$ longer
      (equation~\ref{eq:exptime}) than the average values shown here.}
    \label{fig:exptime}
  \end{figure}
}

\newcommand{\figFsupower}{
  \begin{figure}[htbp]
    \centering
    \plotone{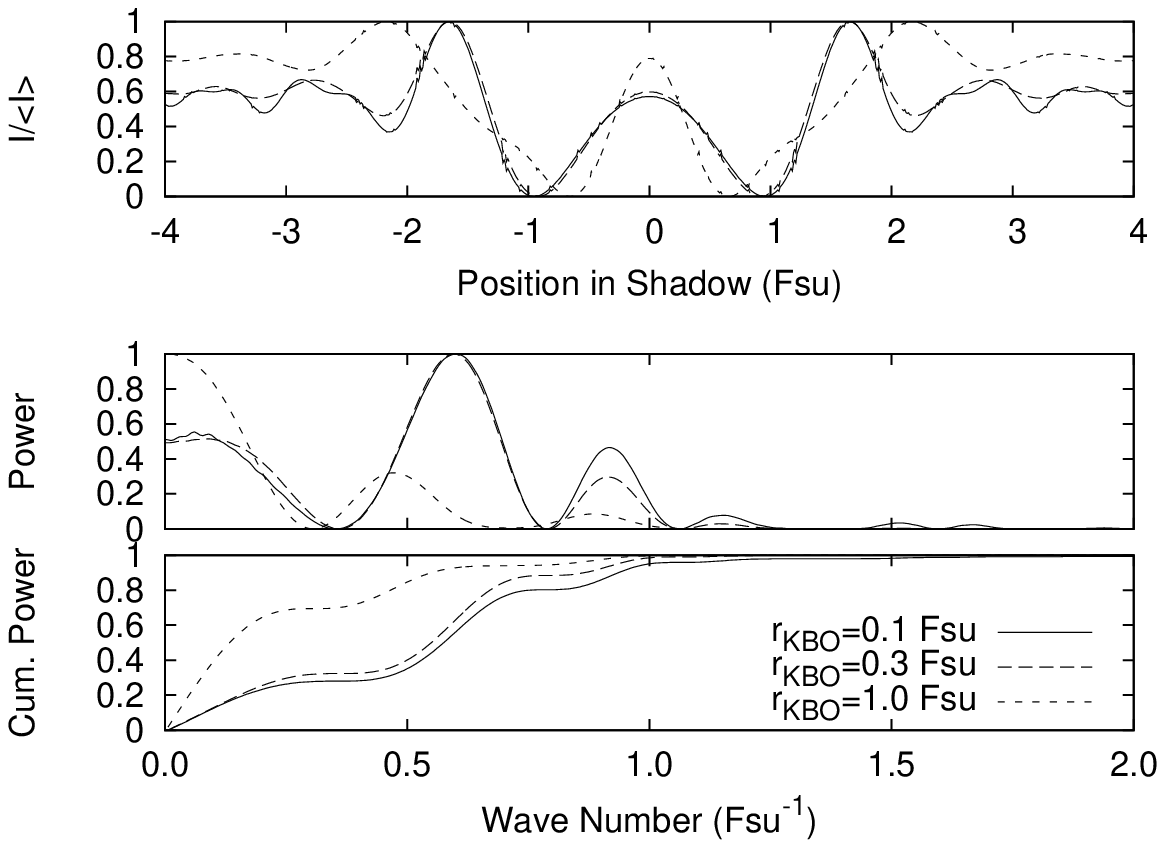}
    \caption{Diffraction-dominated occultation profiles (top) and the
      corresponding power spectra (middle), and cumulative power
      (bottom).  The cumulative power reaches 95\% at a frequency
      $\fny<1\ \invFsu$, and Nyquist sampling is achieved at a sampling
      rate $f_s=2\ \invFsu$.}
    \label{fig:fsupower}
  \end{figure}
}

\newcommand{\figOccrateVsElong}{
  \begin{figure}[htbp]
    \centering
    \plotone{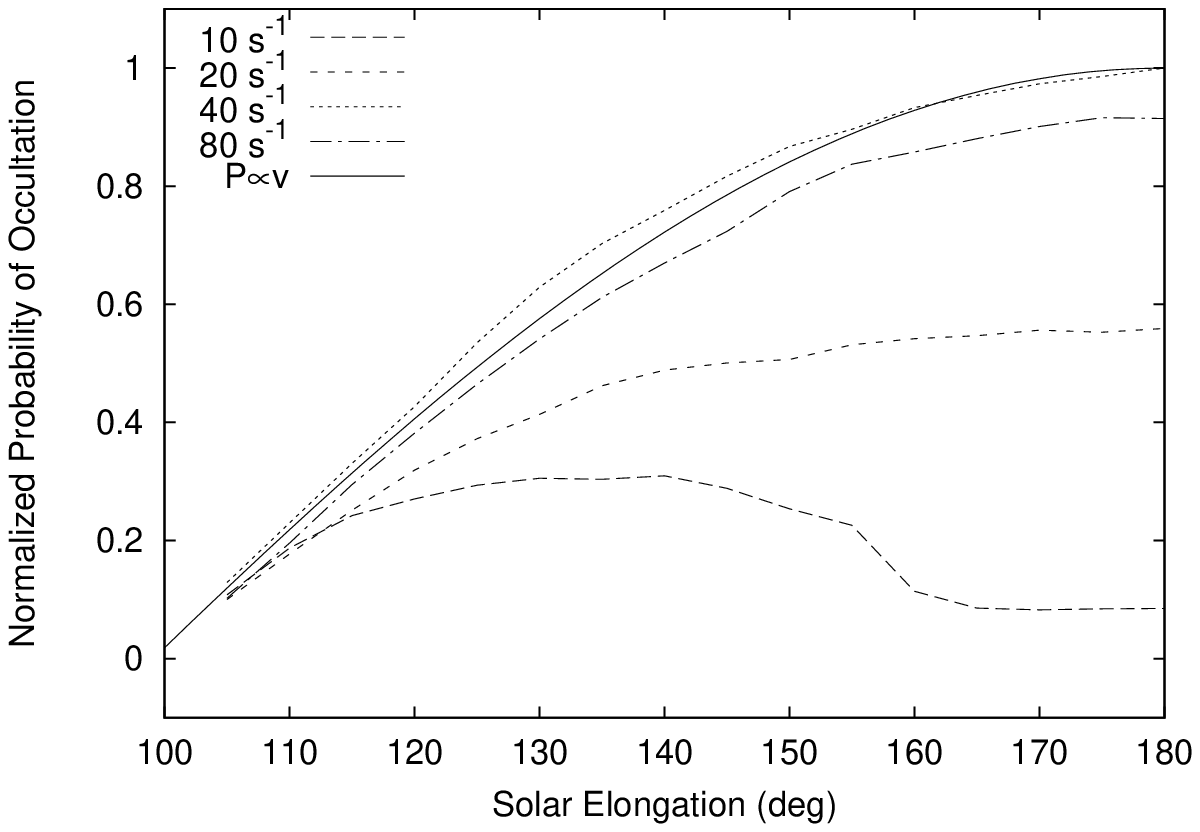}
    \caption
    {The relative detectable occultation rate as a function of solar
      elongation (normalized to an elongation of 180$\arcdeg$) for 10, 20,
      40, and 80 $\invs\ $sampling rates.  The
      reduction of D$_\mathrm{min}$ (Figure~\ref{fig:dmin_vs_elong})
      gives some off-opposition benefit for sub-Nyquist sampling rates
      $\lesssim$40 \invs.  Events are sufficiently sampled at $\gtrsim$40
      \invs, and any such advantage is lost.}
    \label{fig:occrate_vs_elong}
  \end{figure}
}

\newcommand{\figDminVsElong}{
  \begin{figure}[htbp]
    \centering
    \plotone{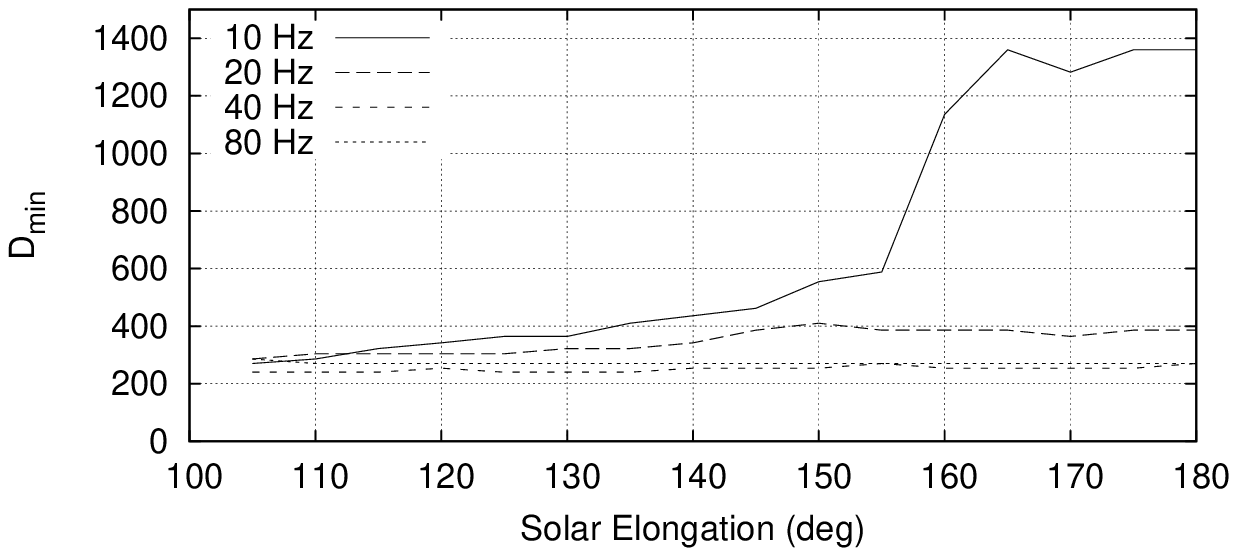}
    \caption
    {The minimum detectable size as a function of solar elongation.
      The improved event sampling due to lower relative velocities
      allows detection of smaller objects for $f_s < 20\ \invs$, but
      has little effect at higher sampling rates.  }
    \label{fig:dmin_vs_elong}
  \end{figure}
}

\newcommand{\figSampRate}{
  \begin{figure}[htbp]
    \centering
    \plotone{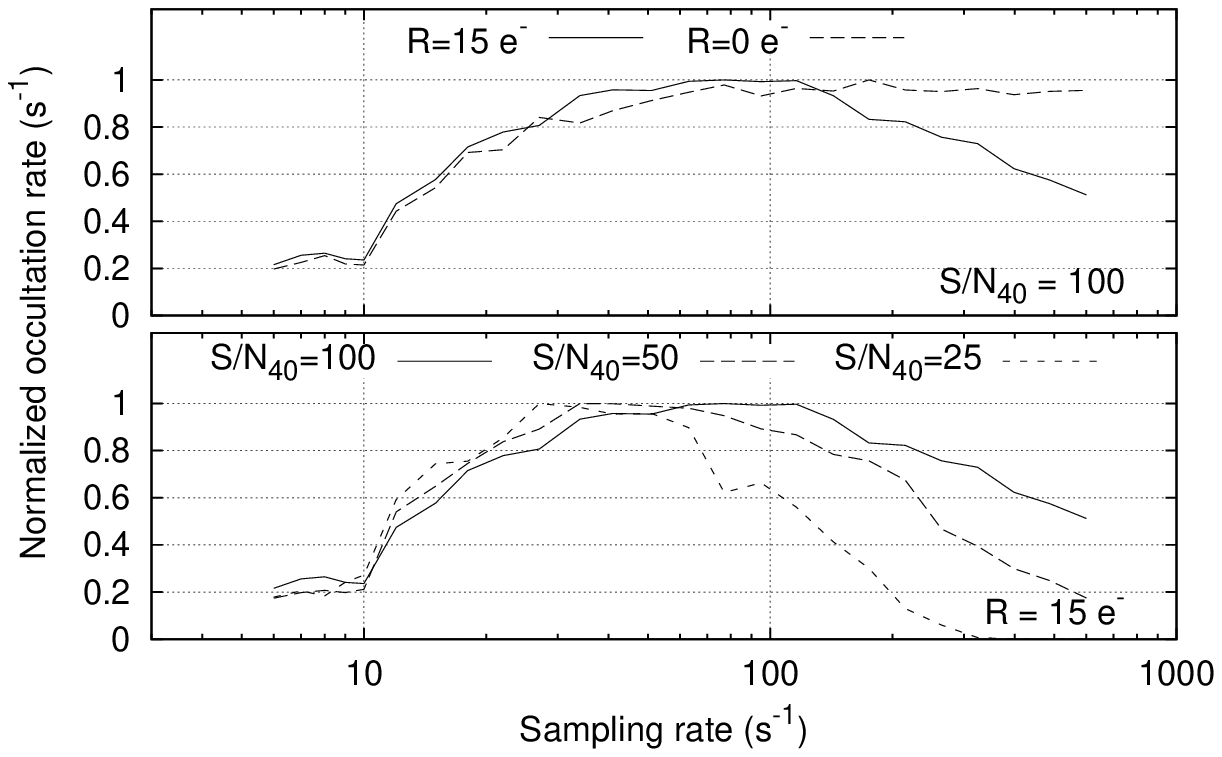}
    \caption
    {The detectable occultation rate (normalized to peak value) as a
      function of the sampling rate.  Both the influence of read noise
      (top) and that of photon noise (bottom) are shown.  In each
      case, detectability peaks at the theoretical $f_s = 40\ \invs$
      rate predicted in Section~\ref{ssec:nyquist}.  The presence of
      read noise reduces the detection probability at high sampling
      rates, and the effect becomes more pronounced at low signal to
      noise.}
    \label{fig:samprate}
  \end{figure}
}

\newcommand{\figNoiseLevel}{
  \begin{figure}[htbp]
    \centering
    \plotone{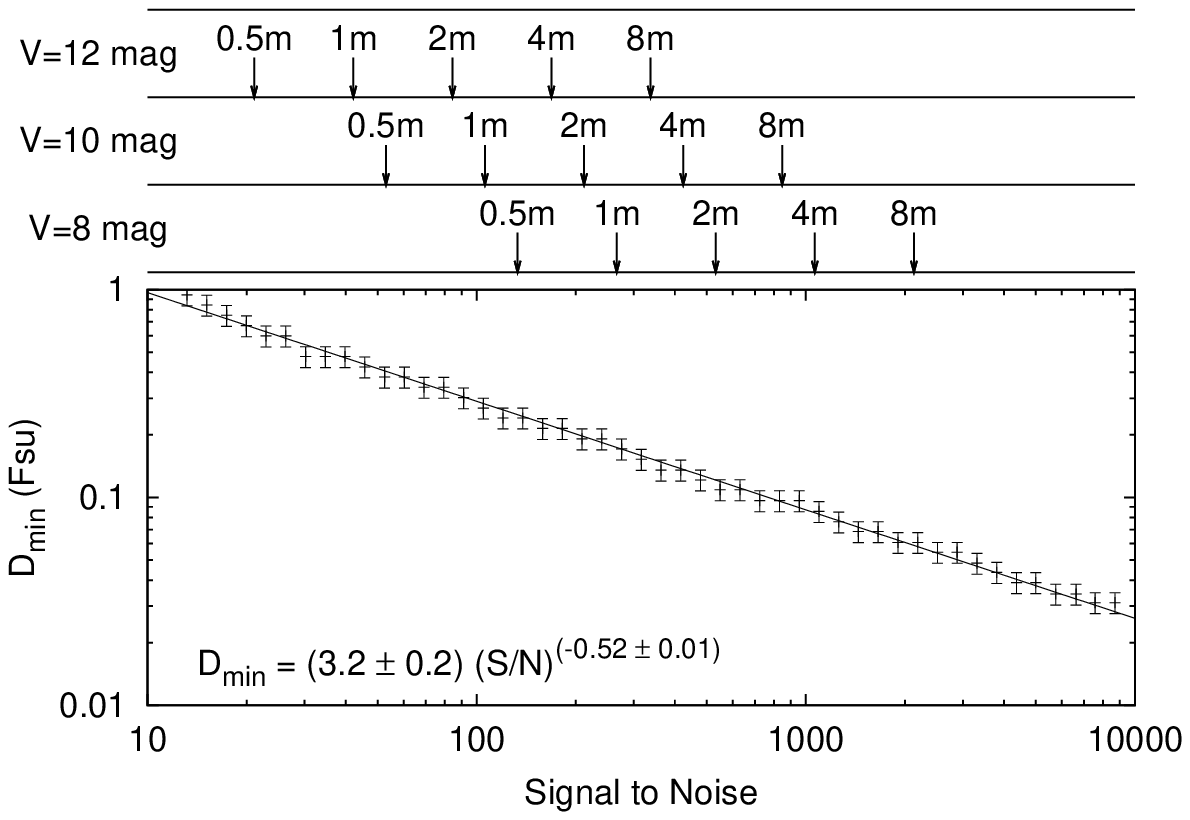}
    \caption
    {The minimum detectable size as a function of the photometric
      signal to noise.  The telescope apertures required to achieve
      the S/N values presented are shown above for V=8, 10, and 12 mag
      target stars. }
    \label{fig:noiselevel}
  \end{figure}
}

\newcommand{\figSDdegen}{
  \begin{figure}[htbp]
    \centering
    \plotone{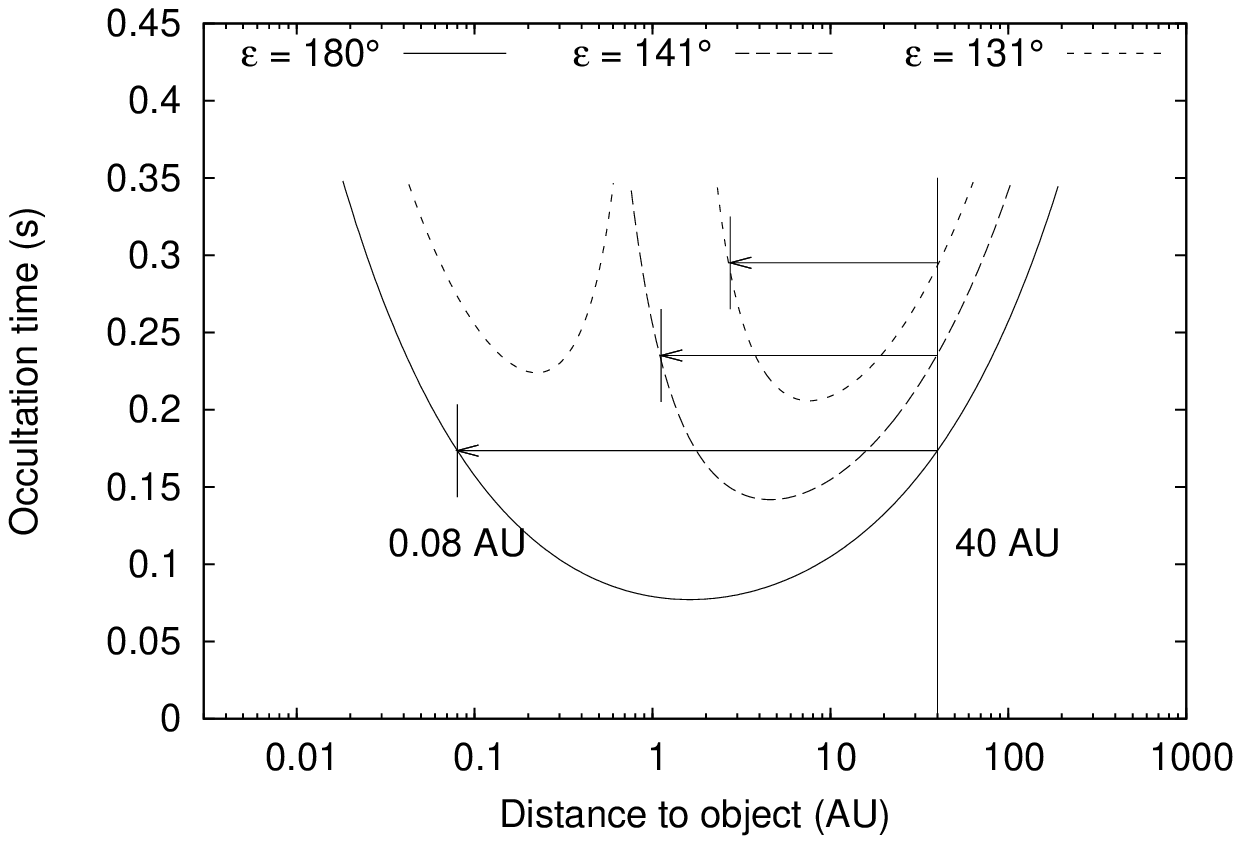}
    \caption
    {The duration of occultation as a function of the distance to the
      occulter.  Different curves represent observations made at
      different solar elongations.  Degeneracy of size/distance occurs
      within the Main Belt for elongations $131\arcdeg \lesssim
      \varepsilon \lesssim 141\arcdeg$.}
    \label{fig:occTime}
  \end{figure}
}

\newcommand{\figSigNorm}{
  \begin{figure}[htbp]
    \centering
    \plotone{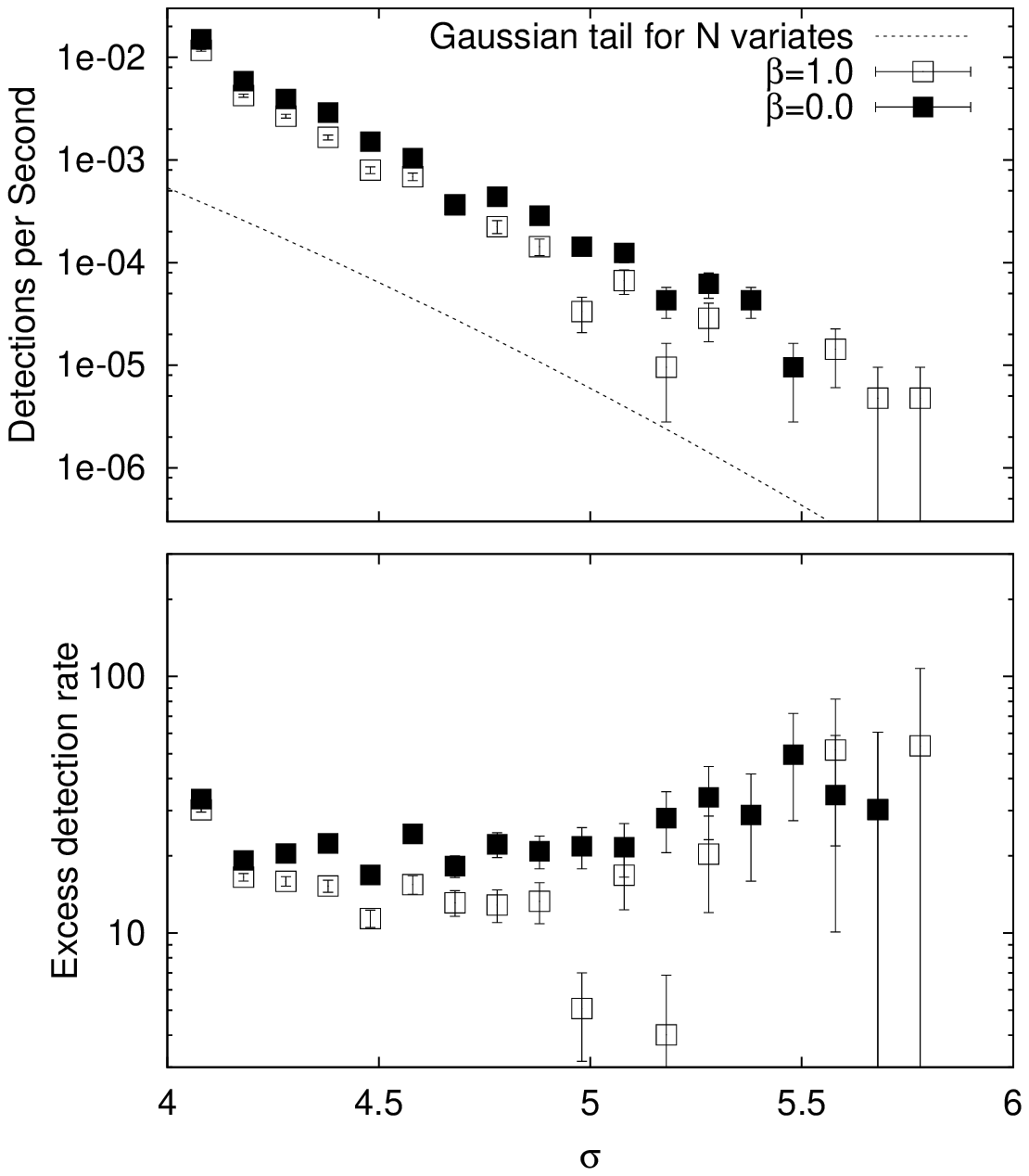}
    \caption
    {The distribution of false positives in units of standard
      deviations shown in events per second (top), and as a fractional
      excess above the rate expected for Gaussian variates (bottom).
      The use of multiple detection kernels causes the number of false
      positives to be $\sim10-20\times$ higher than the rate expected
      for Gaussian variates.  Changes in the signal to noise do not
      significantly alter the rates, but changing the slope of the
      power spectrum from -1 (1/f noise) to 0 (Poisson noise)
      increased rates by a factor of $\sim2$.  }
    \label{fig:sigmaDistrib}
  \end{figure}
}

\newcommand{\figSigSelect}{
  \begin{figure}[htbp]
    \centering
    \plotone{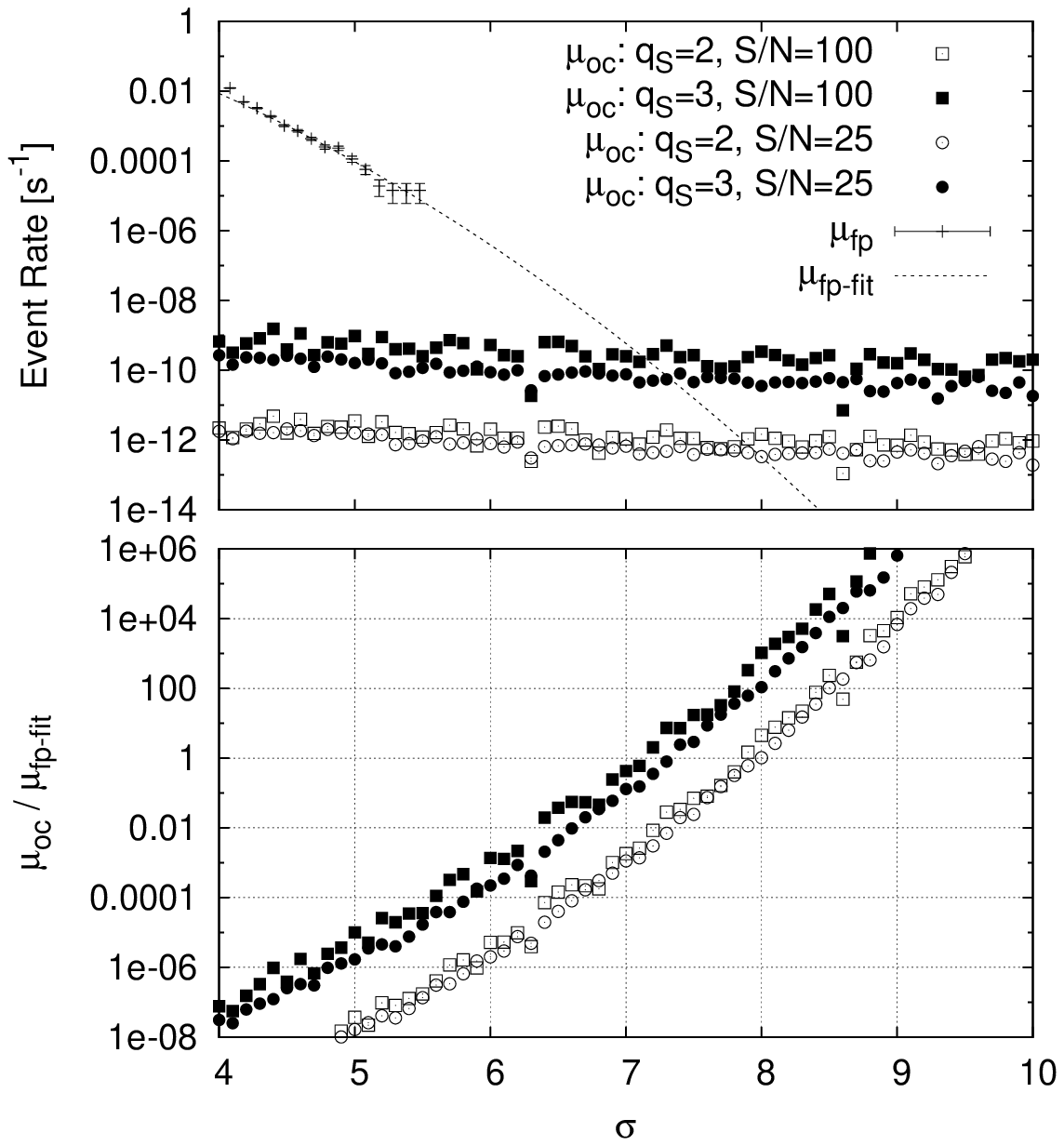}
    \caption
    {The distributions of false and genuine positive occultation
      events in units of standard deviation (top), and the ratios of
      real-to-false positive events (bottom).  False positives
      dominate for significance of $\lesssim8\sigma$.}
    \label{fig:sigselect}
  \end{figure}
}

\ifthenelse{\boolean{astroph}}{
  \newcommand{\middfig}[1]{{#1}}
  \newcommand{\tailfig}[1]{{}}
  
}{
  \newcommand{\tailfig}[1]{{#1}}
  \newcommand{\middfig}[1]{{}} 
}

\shorttitle{KBO Occultations: Rates and Survey Design}
\shortauthors{Bickerton et al.}

\begin{document}

\title{Kuiper Belt Object Occultations: Expected Rates, False Positives, and
  Survey Design}

\author{S.J. Bickerton}
\affil{Department of Astrophysical Sciences, Princeton University,
  Princeton, NJ 08544}
\email{bick@astro.princeton.edu}

\author{D.L. Welch}
\affil{Department of Physics \& Astronomy, McMaster University,
  Hamilton, ON L8S 4M1}
\email{welch@physics.mcmaster.ca}

\and

\author{JJ. Kavelaars}
\affil{Herzberg Institute of Astrophysics, Victoria, BC V9E 2E7}
\email{JJ.Kavelaars@nrc-cnrc.gc.ca}

\begin{abstract}
  A novel method of generating artificial scintillation noise is
  developed and used to evaluate occultation rates and false positive
  rates for surveys probing the Kuiper Belt with the method of
  serendipitous stellar occultations.  A thorough examination of
  survey design shows that: (1) diffraction-dominated occultations are
  critically (Nyquist) sampled at a rate of 2 \invFsu, corresponding
  to 40 \invs for objects at 40 AU, (2) occultation detection rates
  are maximized when targets are observed at solar opposition, (3)
  Main Belt Asteroids will produce occultations lightcurves identical
  to those of Kuiper Belt Objects if target stars are observed at
  solar elongations of: $116\arcdeg \lesssim \varepsilon \lesssim
  125\arcdeg$, or $131\arcdeg \lesssim\varepsilon\lesssim 141\arcdeg$,
  and (4) genuine KBO occultations are likely to be so rare that a
  detection threshold of $\gtrsim7-8\sigma$ should be adopted to
  ensure that viable candidate events can be disentangled from false
  positives.
\end{abstract}


%  Introduction Chapter
%
\section{Introduction}
\label{sec:intro}

A method of detecting km-sized Kuiper Belt Objects (KBOs) is that of
serendipitous stellar occultations (SSO).  Km-sized KBOs are too faint
to be observed directly, but can, in theory, be detected
indirectly. The passage of a KBO through the line of sight to a
background star will produce a characteristic perturbation in the
observed stellar photometric time series, serendipitously revealing
the presence of the occulter.  The method was originally proposed by
\citet{bailey76}, and has been developed extensively
\citep{dyson92,brown97,roques00,cooray03a,cooray03b,gaudi04,nihei07}.
Both optical \citep[][hereafter
BKW]{roques04,roques06,zhang08,bickerton08}, and x-ray
\citep{chang06,jones06} occultation searches have been used to probe
the population of the trans-Neptunian region.  In this paper, we
evaluate a variety of issues which are central to planning, and
performing a KBO occultation survey.  The methods we've used in our
analyses are presented in Section~\ref{sec:methods}, and results for
the issues examined are presented in Section~\ref{sec:examination}.

% 1/f scint noise
Much of the work presented herein is based on the planting and
recovery of occultation profiles in synthetic photometric time series.
Great care must be taken to reproduce the atmospheric scintillation
properties of real photometric data, and we begin in
Section~\ref{ssec:scintillation} by developing a method of generating
artificial time series to reproduce the properties of real time series
in both the time domain and the frequency domain.

% event rate estimates
The SSO method was examined in the context of the Kuiper Belt
\citep[][hereafter R\&M]{roques00} before a break in the KBO size
distribution had been identified \citep{bernstein04} or confirmed
\citep{fuentes08,fraser09}.  In Section~\ref{ssec:occrates}, we present
updated KBO occultation rate estimates based on a broken power-law
size distribution for the KBOs, and with consideration for the
detectability of diffraction-dominated occultation events.

% nyquist sampling
Photometric time series in KBO occultation surveys have been obtained
at a variety of sampling frequencies: 5 \invs \citep{zhang08}, 40
\invs (BKW), and 45 \invs \citep{roques06}.  In
Section~\ref{ssec:nyquist}, we show how Fourier methods can be used to
establish a critical (Nyquist) sampling rate for occultation work.
Numerical tests of the theory are presented in
Section~\ref{ssec:sample}.

% elong tests
The probability an occultation will occur is proportional to the
perpendicular-projected velocity of the occulter with respect to the
Earth.  For the Kuiper Belt, the motion is retrograde and the velocity
is maximized by conducting observations at solar opposition (R\&M).  In
Section~\ref{ssec:elong}, we consider the possibility that
off-opposition observations could improve the detectability of
smaller, and potentially more numerous objects.

% s/n and the event rate
It is often useful to know the diameter of the smallest detectable
object for data with a given signal to noise.  In
Section~\ref{ssec:snr}, we show how the signal to noise ratio,
hereafter S/N, determines the minimum detectable size.

% degeneracy with MBAs
Interest in small body occultations has focused largely on the Kuiper
Belt, but other near-field objects are capable of producing
occultations.  The size/distance degeneracy for occultation events has
been assessed by \citet{cooray03a}, with a discussion on how it may be
broken for diffraction-dominated events.  In Section~\ref{ssec:degen},
we evaluate the size/distance degeneracy further to determine the
size/distance an occulter must have to produce an event identical to a
KBO occultation.  We then consider the possibility of degeneracy with
the Main Belt Asteroids (MBAs).

% fp tests
We close in Section~\ref{ssec:falsepos} with an evaluation of how
atmospheric scintillation influences the false positive occultation
rate.  We then compare false positive rates to the expected rates for
real occultations to determine an appropriate detection threshold.

Results are discussed and summarized in Sections~\ref{sec:discussion},
and~\ref{sec:summary}, respectively.

\section{Methods of Analysis}
\label{sec:methods}

Optimization of the observing parameters in occultation work requires
a metric by which the different parameters can be compared.  We have
chosen to measure how successfully an event can be detected with a
`plant and recover' strategy.  There are three contributing components
in this approach: a simulated occultation lightcurve, a simulated
photometric time series, and a detection algorithm.

\subsection{Simulating Occultation Lightcurves}
\label{ssec:kernel}

Diffraction effects become significant in an occultation when the
occulting object has a radius smaller than the Fresnel scale unit, $1\
\Fsu = (\lambda d/2)^{1/2}$, where $\lambda$ is the wavelength of
observation, and $d$ is the distance to the occulter.  At 40 AU the
Fresnel scale for 550 nm light is 1285 m.  A successful occultation
lightcurve model for a km-sized KBO must therefore take into account
diffraction effects \citep{roques87,nihei07,bickerton08}.  We have
adopted a lightcurve modeling method described by \citet{roques87} in
which a circular occulting mask is assembled by placing progressively
smaller rectangles at the periphery of the mask. Five ``orders'' of
these rectangles have been used for all occultation lightcurve models
in this work.

All models were numerically integrated over a stellar disk with a
projected size appropriate for the distance to, and spectral type of,
the star being considered.  Similarly, models were numerically
integrated over a 400-700 nm passband.

% Scintillation subsection
%
\subsection{Simulating Photometric Time Series}
\label{ssec:scintillation}

The evaluation of false positive rates (Section~\ref{ssec:falsepos})
requires that realistic artificial time series be analyzed.
\citet{dravins97a} showed that scintillation in a photometric time
series produces a power spectrum which decreases as a function of
frequency.  An artificial power spectrum can be generated and tailored
to reproduce the properties of an observed one, and this artificial
power spectrum can then be inverse-Fourier transformed to produce a
time series which has the same statistical properties -- both in time
and frequency -- as the original `real' data set.

\subsubsection{The Power Spectra of Photometric Time Series}
\label{sssec:powerspectrum}

Discrete Fourier transforms \citep[see, for example][]{bracewell86} of
the photometric time series described in BKW were
computed with a fast Fourier transform (FFT) algorithm.  FFT
algorithms are most efficient for time series composed of $N=2^n$
points, where $n$ is an integer, and the time series were truncated to
have the largest number of points which satisfied this requirement.
The real and imaginary frequency components for a 409.5 s time series
(2$^{14}$=16384 points at 40 s$^{-1}$) are shown in the upper panels
of Figure~\ref{fig:FTdistrib}.  The amplitude of the components
decreases as a function of frequency.  Each frequency component was
normalized by the standard deviation of the values within $\pm0.5
\invs$, and the normalized frequency component distributions are
presented in the lower panels of Figure~\ref{fig:FTdistrib}.  The
frequency components are consistent with Gaussian variates at 95\%
confidence (based on reduced $\chi^2$ tests).

\middfig{\figFTnoise}

The power spectrum corresponding to Figure~\ref{fig:FTdistrib} is
shown in Figure~\ref{fig:powerspectrum} and can be seen to obey $\log
|F|^2 \propto \beta \log f$, or $|F|^2 \propto f^{\beta}$, where $|F|$
is the norm of the frequency components, $f$ is the frequency, and
$\beta$ is the power law slope.  The power spectrum approaches a value
of zero as the highest frequency is approached, indicating that
aliasing effects are likely to have been minimized \citep{press92}.
In observed photometric time series, no relationship between phase and
frequency was expected or observed.

\middfig{\figPowerSpec}

The photometric time series reported in BKW (56 separate 7-25 minute
segments) were Fourier transformed to produce power spectra and the
power law slopes were measured with linear least squares for
frequencies: $0.1\ \invs < f < 20\ \invs$.  The measured slopes were
normally distributed with $\overline{\beta}=-1.0$, and
$\sigma_{\beta}=0.3$.  The value of $\beta$ was not observed to vary
as a function of frequency.  As the time series were found to contain
no detectable occultation events, the structure in these power spectra
reflects the properties of the noise.

\subsubsection{Generating Artificial Time Series}
\label{sssec:generating}

Generation of artificial time series began with the generation of
artificial frequency components $\Re H(f_n)$ and $\Im H(f_n)$.  The
frequency components were inverse Fourier transformed to generate the
time series.

\begin{equation}
  h_k = \frac{1}{N} \sum_{n=0}^{N-1} H_n e^{-2\pi ikn/N}.
\end{equation}

At a frequency, $f_n$, the real and imaginary components were assigned
with phases, $\theta_n$, amplitudes, $a_n$, and a normalizing
constant, $A$, to produce a power spectrum with slope $\beta$:

\begin{eqnarray}
  \label{eq:real}
  \Re\ H(f_n) &=& f_n^{\beta/2} A\ a_n \cos(\theta_n), \quad \mbox{and}\\
  \label{eq:imag}
  \Im\ H(f_n) &=& f_n^{\beta/2} A\ a_n \sin(\theta_n).
\end{eqnarray}

The phase angles were drawn from a uniform random distribution with
$-\pi \leq \theta < \pi$, and the amplitudes, $a_n$, were drawn from a
Gaussian distribution with $\overline{a}$=0, and $\sigma_a=1$.

To produce a time series which is entirely real-valued, the property of
conjugate symmetry \citep[see, for example][]{bracewell86} requires
that the spectral components for the negative frequencies are the
complex conjugates of their positive counterparts, $F(f_{N-n}) = F^\star(f_n)$.

From the definition of the Fourier transform, $H(0)$ is determined by
the mean, $\overline{h}$, of the time series being produced:

\begin{eqnarray}
  \label{eq:parseval}
  H_k &=& \sum_{n=0}^{N-1} h_n\ e^{2\pi ikn/N},\\
  \label{eq:H0}
  H_0 &=& \sum_{n=0}^{N-1} h_n = N\overline{h},
\end{eqnarray}

\noindent where $N$ is the number of points.

The normalization constant, $A$, was chosen to produce the desired
noise level, $\sigma_h$, in the time series (see Appendix~\ref{app:Anorm}):

\begin{equation}
  \label{eq:Anorm}
  A = N \sigma_h \left[ \sum_{n=0}^{N-1} f_n^\beta a_n^2 \right]^{-1/2}.
\end{equation}

Frequency and time components for a 128-point artificial data set are
presented in Figure~\ref{fig:FTfake} to illustrate the
method\footnote{The code used to compute the algorithm is freely
  available from SJB.}.

\middfig{\figFTfake}

The variates of a true photometric time series are log-normally
distributed; but for variances of $(\sigma_h/\overline{h})^2 \lesssim
0.1$, the distribution is indistinguishable from a Gaussian
\citep{dravins97a}.  If the chosen noise level ($\sigma_h$) is too
high, the method will produce some negative variates, inconsistent
with photometric measurements.  This method should only be considered
valid for noise levels of a few percent, $\sigma_h/\overline{h}
\lesssim 0.1$.  

One notable difference between observed and artificial photometric
time series is the variability in the noise level.  Atmospheric
transparency can change on time scales of a few minutes during an
observation, and airmass changes can be present in longer
observations.  The associated changes in the flux level can be
normalized by smoothing, but normalization of the noise level (eg. by
a moving standard deviation) would distort the amplitude of an
occultation signal.  To model a time series containing significant
changes in the overall photometric noise level, the method above can
be extended by modulating the amplitude of the artificial time series.

If photometric noise levels are below a few percent, and remain stable
throughout a time series, an artificial time series data can be
generated to simulate scintillation effects consistent with the real
photometric measurements.  The artificial time series will have the
same statistical properties as the real data in both the time domain,
and the frequency domain.

\subsection{The Detection Algorithm}
\label{ssec:detection}

There are three different detection methods used to identify
occultation events in photometric time series: the variability index
\citep{roques06}, the cross-correlation of template lightcurves
\citep{bickerton08}, and the rank probability method \citep{zhang08}.  The
cross-correlation of template occultation lightcurves has been used in
this work.

Artificial time series, $t_i$, were generated for a given sampling
rate (Section~\ref{ssec:scintillation}), and template occultation
lightcurves were constructed (Section~\ref{ssec:kernel}). The
occultation lightcurves were discretely sampled at the same rate as
the time series to form cross-correlation test kernels, $k_i$, and the
kernels were then cross-correlated with the time series:

\begin{equation}
  \label{eq:xcor}
  (t \star k)_j \equiv \sum_i t_i k_{i+j}.
\end{equation}

A positive peak in $(t \star k)_j$ indicates that the structure in the
time series, $t$, is similar to that of the test kernel, $k$, when $k$
is offset at position $j$.  The detection threshold was set to
$>+8\sigma_{t\star k}$, the limit at which real occultation events are
expected to occur at a higher rate than false positives (see
Section~\ref{ssec:falsepos}).

A suite of test kernels was selected to provide coverage over
the distances, occulter sizes, and occultation impact parameters of
interest in the cases being examined.

\section{Occultation Survey Parameters}
\label{sec:examination}

Using the methods described in Section~\ref{sec:methods}, we now
examine several observing parameters which are of importance in
occultation work.  Numerical simulations have been used to determine
how these parameters influence the rate at which detectable
occultation events occur.  Whenever possible, optimal values are
established.

We begin by determining the expected KBO occultation rate, and proceed
to examine how it is effected by changes in the various observing
parameters.

% Occultation rate vs size
\subsection{Expected KBO Occultation Rates}
\label{ssec:occrates}

Existing estimates of the KBO occultation rate (R\&M) were
based on a single power law size distribution, which has since has
been succeeded by a broken power law model
\citep{kenyon04,pan05,bernstein04,fuentes08,fraser09}.  Here, we
evaluate the occultation rate expected for a KBO sky surface density
based on the more recent broken power law model.

An occultation at distance, $d$ [m], with maximum detectable impact
parameter, $b$ [m], projected velocity, $v$ [m \invs] (projected onto
the plane perpendicular to the line of sight), and the sky surface
density, $\Sigma$ [\pdegsqrd]; will occur at rate:

\begin{equation}
  \label{eq:2bvs}
  \mu = 2bv\Sigma \left(\frac{180}{\pi d}\right)^2 \quad [\invs].
\end{equation}

The Poisson probability of observing $x$ occultations in time $t$ is:
$P(x) = [(\mu t)^x/x!]\ e^{-\mu t}$, and the expected time between
occultations is (BKW):

\begin{equation}
  \label{eq:exptime}
  t_{\mathrm{exp}} = \frac{\ln |1-P(x>0)|}{-\mu}.
\end{equation}

\noindent Values of $b$, $v$, and $\Sigma$ are determined as follows.

The relative perpendicular velocity of an occulter in an uninclined
circular orbit is \citep{liang02}:

\begin{equation}
  \label{eq:vperp}
  v = \left(\frac{\mathrm{G}\Msun}{\rearth} \right)^{\frac{1}{2}}
  \left[
    \left[  \frac{\rearth}{\rocc}  
      \left( 1 -  \left( \frac{\rearth}{\rocc} \right)^2
        \sin^2(\varepsilon)\right)
    \right]^{\frac{1}{2}} +
    \cos(\varepsilon)
  \right],
\end{equation}

\noindent where $r_{\oplus,\mathrm{o}}$ are the orbital radii of the
Earth and occulter, $\varepsilon$ is the solar elongation of the
target star, and $G$ and \Msun are the gravitational constant and solar
mass.

We model the sky surface density, $\Sigma$, with the broken power law
cumulative size distribution described by \citet[][eqs. 12 \&
13]{gladman01}.  Known values have been expressed in a single constant
of proportionality, $Q_1$.

\begin{equation}
  \label{eq:NoverS}
  \Sigma = Q_1 \ D_k^{\ \qS-4.8}\ D_0^{1-\qS} \ \ [\pdegsqrd],
  \quad Q_1 = 3.2\times10^8.
\end{equation}

\noindent Here, $D_0$ [km] is the diameter of the smallest KBO
considered, and $D_k$ [km] is the diameter at which the power law
breaks from the large object slope, $\qL=4.8$
\citep{fraser09,fuentes08}, to the small object collisional
equilibrium slope $\qS$.

The widths of occultation shadows (the `$2b$' term in equation~\ref{eq:exptime})
take different values for smaller objects in the diffraction dominated
regime, than for larger objects, for which such effects can safely be
ignored.  \citet{nihei07} define shadow width by the diameter of the
Airy ring ($\sim 2\sqrt{3}$ Fsu = $(6\lambda d)^{1/2}$), with a smooth
transition between the two limiting cases:

\begin{equation}
  \label{eq:width}
  W = \left[ \left( (6 \lambda d)^{1/2} \right)^{3/2} + D^{3/2} \right]^{2/3} + 2R_{\star}.
\end{equation}

Simulations have shown this width is a reasonable estimate of the
detectable width of an occultation (see for example, BKW, Figures 7 \&
12).  The cumulative surface density, $\Sigma$, includes objects
larger than $D$, and a mean width, $\overline{W}$, was used to
represent the average occulter.

\begin{equation}
  \label{eq:Wbar}
  2b = \overline{W} \approx 
  \left[\left((6\lambda d)^{1/2}\right)^{3/2} + \overline{D}^{3/2}\right]^{2/3} + 2R_\star,
  \end{equation}

\noindent where (derivation in Appendix~\ref{app:Dbar}):

\begin{eqnarray}
  \label{eq:meanwidth}
  \overline{D}(D>D_0) &=& \frac{1}{N} \int_{D_0}^{\infty} \frac{\dif N(D)}{\dif D} D\ \dif D\\
  \label{eq:meanwidthfinal}
  &=& \left\{
    \begin{array}{ll}
      \left[ \left( \frac{1-\qL}{2-\qL} - \frac{1-\qS}{2-\qS} \right) \left( \frac{D_0}{D_k}\right)^{\qS-2} \right.&\\ \nonumber
      \left. \qquad + \quad \frac{1-\qS}{2-\qS} \right] D_0.& (\qS\ne2)\\ \nonumber
      \left[ \frac{1-\qL}{2-\qL} +\ln\left(\frac{D_k}{D_0}\right) \right] D_0 & (\qS=2)
    \end{array}
  \right.
\end{eqnarray}

\noindent The fractional exponents 2/3 and 3/2 in
equation~\ref{eq:width} provide a smooth transition between terms, and
the substitution of $\overline{D}$ to obtain $\overline{W}$ gives a
reliable first order approximation\footnote{As $W$ is not a linear
  function of $D$, $\overline{D}$ cannot be directly substituted for
  $D$ to obtain $\overline{W}$ (as in equation~\ref{eq:Wbar}).
  However, to a first-order approximation, $W$ {\itshape is} a linear
  function of $D$, and the substitution is valid as an
  approximation.}.

Times required to observe one or more occultations by $D > D_0$
objects are shown in Figure~\ref{fig:exptime} for optical
($\lambda=550$nm) and X-ray ($\lambda=0.4$nm) light (the bands
observed in occultation searches to date).  The expected average times
between events are shown (equation~\ref{eq:exptime} with
$P(x>0)$=0.68, or 68\% confidence for one or more events).  Curves
were computed for $\qS$=2.0 \citep{fuentes08,fraser09}, and 3.0. Other
parameter values were: $R_{\star}$=0 (point source), distance $d$=40
AU, solar elongation $\varepsilon$=180$^\circ$ (opposition),
$D_k$=50km \citep{bernstein04}, and $\qL$=4.8 \citep{fraser09}.

\middfig{\figOccrates}

The 95\% confidence times may be more appropriate estimates for
programs seeking to detect a single event.  From
equation~\ref{eq:exptime}, these are $-\ln|0.05| \approx 3\times$
longer.

The expected occultation rates are similar for X-ray and optical
observations.  Although the X-ray Fresnel scale is small enough to
offer access to the more-numerous 10m-sized objects, there is only a
single bright ecliptic X-ray target (Sco X-1).  The relative abundance
of suitable optical targets (eg. M35, with 100s of upper main sequence
stars) makes them a better choice for a large scale KBO occultation
survey.

\subsection{Nyquist Sampling for a Diffraction-Dominated Occultation Event}
\label{ssec:nyquist}

A measured signal should be sampled at a frequency, $f_s$, which is
2$\times$ the Nyquist frequency, \fny (highest frequency to be
represented). A diffraction-dominated occultation event contains some
power at all frequencies, and the Nyquist frequency is defined here to
be the sampling rate for which 95\% of the cumulative power is present
in frequencies below \fny.  This limit is, and must be, somewhat
arbitrary.  The goal was to identify an upper limit for which the
overwhelming majority of the occultation signal strength would be
represented in the sampled measurements.  The validity of this
95\%-power limit is supported by numerical tests which are presented
in Section~\ref{ssec:sample}.

Power-spectra were computed for the lightcurves of \rKBO=0.1, 0.3,
and 1.0 Fsu occulters.  The Fresnel-scale unit is a function of
wavelength, $\lambda$, and integration over a passband was simulated
by averaging lightcurves within $\rKBO\pm$15\% (eg. $0.085 < \rKBO <
0.115$ for $\rKBO=0.1$ Fsu).  This $\rKBO\pm15\%$ range simulates
integration over a $\lambda\pm27\%$ passband, and was chosen to
represent the 400-700 nm passband of visible light.  The lightcurves,
power spectra, and cumulative power spectra are shown in
Figure~\ref{fig:fsupower}.  Conveniently, the cumulative power reaches
95\% at a wave number of $k$=1 $\invFsu$, regardless of $\rKBO$,
corresponding to a Nyquist-sampled sampling rate of $f_s$=2 \invFsu.

\middfig{\figFsupower}

The lack of dependence on $\rKBO$ would not generally be the
case for $\rKBO \ll 1$ Fsu lightcurves observed in monochromatic
light.  A significant portion of the power would then be present in
extended high-frequency ringing.  When a realistic integration over a
passband is considered, the extended ringing cancels and the \fny = 1
Fsu limit is valid, even for the $\rKBO = 0.1$ Fsu example shown in
Figure~\ref{fig:fsupower}.

For visible-light observations ($\lambda$=400-700 nm) of KBOs at
distance $d$=40 AU, an occultation with a projected velocity of $v$=26
\kmps (ie. at solar opposition) would be Nyquist sampled at:

\begin{eqnarray}
  \label{eq:fNyquist}
  f_s &=& 2\ \invFsu \cdot \left(\frac{v}{(\lambda d / 2)^{1/2}}\right)\ \Fsups\\
  &=& 40\ \invs. \nonumber
\end{eqnarray}

\noindent Beyond 40 AU, the projected velocity asymptotically
approaches 30 \kmps, and the sampling rate scales as $f_s \propto
d^{-1/2}$ until the projected size of the target star is comparable to
the Fresnel scale.

% Occultation rate vs elong
\subsection{Solar Elongation of the Target Star}
\label{ssec:elong}

Occultations observed off-opposition have lower relative velocities
perpendicular to the line of sight, and are better-sampled events.
But, they are less probable events as the occultation rate increases
linearly with velocity, $\mu = 2bv\Sigma$ (equation~\ref{eq:2bvs}).
Here, we determine how the combination of increased detectability and
decreased event probability effect the observed occultation event rate
as a function of solar elongation.

% made fresfiles 50m - 250m (5m incr) for 40 AU
Diffraction dominated occultation profiles were generated for 50
logarithmically-spaced circular occulters with sizes 100 m $< \rKBO <$
2000 m, at 40 AU, observed in optical light (400--700 nm).  The
circular occulting masks were assembled out of rectangular masks
(Roques et al.~\citeyear{roques87}; BKW), and target stars were
assumed to be point sources.

% made timeseries for 20,40,80Hz
Artificial time series were generated with 1\% 1/f noise (S/N = 100),
as described in Section~\ref{ssec:scintillation}.  Tests were also
performed with noise levels of 2\% and 4\%, and gave similar results.
Time series had 65536 points, and were tested with sampling rates of 10~\invs,
20~\invs, 40~\invs, and 80~\invs (above and below the Nyquist limit
described in Section~\ref{ssec:nyquist}).

% added 100 events to each
Relative velocities were computed at solar elongations between
opposition and the stationary point (100 - 180$^\circ$ in increments
of 5$^\circ$), and the template diffraction patterns were scaled to
produce templates corresponding to each velocity.  The scaled
templates were sampled at the rates listed above and 10 simulated
events were planted in the time series.  This was repeated for 30
evenly-spaced occultation impact parameters extending to 8$\times$ the
Fresnel scale.  The planting process is described in BKW.

% fitted limiting size r_min
The Cross-correlation detection algorithm was used to attempt
detection of planted events (Section~\ref{ssec:detection}), and the
fraction of recovered events transitioned between $\sim$1 for large,
easily detected events, and 0 for events well below the detection
threshold.  The occultation rate was integrated by summing
contributions from each $\delta r$ radius bin: $\delta\mu = 2
b_{\mathrm{max}}(r) v \delta\Sigma(r,\qS)$, where $b_{\mathrm{max}}$
was the maximum impact parameter detected, and $\delta\Sigma(r,\qS)$
was the sky surface density of objects in the size range $r < \rKBO <
r + \delta r$, for a size distribution having a small-object slope of
$\qS$.  Slopes of \qS=2, and 3 were tested.  The integral was
truncated to exclude objects with $\rKBO>2$ km as the sky surface
density of objects larger than this is comparatively extremely small
for the size distribution slopes tested.  The occultation rates were
normalized to the solar opposition value and are shown in
Figure~\ref{fig:occrate_vs_elong}. Results of the $\qS$=3 tests were
similar those for \qS=2, and are not presented in
Figure~\ref{fig:occrate_vs_elong}.

\middfig{\figOccrateVsElong}

The limiting size, $D_{\mathrm{min}}$, was defined to be the diameter
at which the shadow width, $2b$, reached 1 Fsu.  This was found to be
a robust measure of the diameter at which an object was too small to be
detected.  The minimum detectable sizes versus solar elongation are
shown in Figure~\ref{fig:dmin_vs_elong}.  They were found to decrease
as a function of solar elongation for the sub-Nyquist samplings
(10~\invs and 20~\invs), but did not change significantly at the
higher sampling rates.

\middfig{\figDminVsElong}

The improved off-opposition detectability of the more numerous smaller
objects in the 10~\invs and 20~\invs tests suggests that detection
rates in such an observing program would not decrease in proportion to
$v$.  Above the Nyquist sampling rate, the detection rate decreases in
proportion to $v$, and no improvement in detectability is achieved by
working off-opposition.

% Occultation rate vs sample rate
\subsection{Sampling Effects}
\label{ssec:sample}

The Nyquist limit presented in Section~\ref{ssec:nyquist} indicates
that the theoretical optimum sampling rate is $f_s$=2 \invFsu.  Here
we examine the detection rate for occultation events observed by a
camera operating at different sampling rates.

Occultation profiles and artificial time series were generated as
described in Section~\ref{ssec:elong}.  Our $1/f$ noise model was only
tested for sampling up to 40 \invs (the sampling rate of our observed
time series).  The $1/f$ property of scintillation must ultimately
transition to (white) photon noise at $\gtrsim 100 \invs$
\citep{dravins97a}. Our time series were therefore simulated with
Poisson variates (white noise) to overcome this limitation in our
understanding of the noise beyond 40 \invs.  Though each test was
performed with Poisson variates, it was repeated with the $1/f$ noise
model for frequencies up to $f=40$ \invs, and results were found to be
consistent with the Poisson results for the range of frequencies
tested.  This simulation represents space-based observations very
well, and is representative of ground-based observations up to $f=40$
\invs.

Constant values of D=40 AU (ie. KBOs), and \qS=2.0 were used, and the
sampling rate, $f$, of the camera was varied in logarithmic spacings
$6\ \invs < f < 600\ \invs$.  The tests were bench-marked with respect
to a sampling-specific signal to noise, S/N$_f$.  A bench mark sampling
rate of $f=40\ \invs$ was used, and signal to noise ratios of
S/N$_{40}$=100, 50, 25 were tested. At each sampling rate Poisson
variates were drawn to represent fluxes, $I_i(f)$, with mean
$\overline{I}(f)$ (in photo-electrons):

\begin{equation}
  \label{eq:snrf}
  \overline{I}(f)= \left( \frac{40}{f} \right) (\mathrm{S/N}_{40})^2.
\end{equation}

\noindent Target stars for occultation work are generally bright
(V$<$12 mag) and sky noise was not considered.

Simulated occultation events were planted in the data, and random
Gaussian variates, $R_i(f)$, with a mean, $\overline{R}(f)$=0, and
standard deviation, $\sigma_R(f)$, were then added to each point in
the time series to simulate read noise.  Read noise is a function of
the clamp-and-sample time for a pixel, $t_{\mathrm{CS}}^{-1/2}$,
\citep{mclean97}.  The value, $\sigma_R(f)$, at each sample rate, $f$,
was normalized to the 40 $\invs$ value, $\sigma_R(40)$:

\begin{equation}
  \label{eq:readnoise}
  \sigma_R(f) = \sigma_R(40) \left( \frac{f}{40} \right)^{1/2}.
\end{equation}

It was assumed that $t_{\mathrm{CS}}$ is linearly related to the full
read-out time for the CCD.  This is consistent with a constant-size
region of a frame transfer device being read out at a rate which makes
full use of the exposure time.  The dead-time associated with frame
transfer is typically 1 ms and was not considered. Values of
$\sigma_R(40)$=0, and 15 $\pe$ were tested.

The photon noise plus read noise signals, $I_i(f)+R_i(f)$, were
searched for the planted occultation events with the cross-correlation
detection algorithm, and the smallest detectable size,
$D_{\mathrm{min}}$, was determined as described in
Section~\ref{ssec:elong}.  The density, $\Sigma$, of occulters larger
than $D_{\mathrm{min}}$ was computed for a small object slope \qS=2
(equation~\ref{eq:NoverS}), and the probability of occultation
was integrated over the KBO radii as described in
Section~\ref{ssec:elong}.

The detection probabilities at different sampling rates are presented
in Figure~\ref{fig:samprate}.  The presence of read noise does not
strongly influence the occultation probability when the photon noise
is low (upper panel), but becomes a significant attenuating factor
at higher photon noise levels (lower panel).  This is not surprising
as the read noise is a more significant component of the total noise
level when photon counts are low.

\middfig{\figSampRate}

\subsection{Signal to Noise}
\label{ssec:snr}

Significant increases in the occultation event rate can be achieved by
using a larger telescope to increase the signal to noise in the
photometric time series.  Here, we evaluate the detectable occultation
rate at different photometric noise levels which are representative of
available observing facilities.

KBO (ie. 40 AU) occultation profiles were planted in simulated 40
\invs (ie. optimally sampled) time series containing $1/f$ noise at
logarithmically-spaced signal to noise levels of $10 < \mathrm{S/N} <
10000$.  Detection was attempted with the cross-correlation algorithm
to determine the minimum detectable size, $D_{\mathrm{min}}$, as
described in Section~\ref{ssec:elong}.  Figure~\ref{fig:noiselevel}
shows $D_{\mathrm{min}}$ versus S/N.

\middfig{\figNoiseLevel}

The minimum diameter, in Fsu, empirically follows the power law: 

\begin{equation}
  \label{eq:dmin}
  D_{\mathrm{min}} = Q_2\ (S/N)^{-\eta},
\end{equation}

\noindent with $Q_2 = 3.2\pm0.2$, and $\eta = 0.52\pm0.01$ (by linear
least-squares fit to $\log D_{\mathrm{min}} = \log Q_2 - \eta \log (S/N)$).  This
relation was found to remain valid when tested at different distances,
but is expected to break down at distances for which the target star
has a projected radius $\gtrsim$1 Fsu. Combining the minimum
detectable size with the cumulative size distribution for the small
occulters (equation~\ref{eq:NoverS}) the sky surface density,
$\Sigma$, can be expressed in terms of S/N:

\begin{equation}
\label{eq:detectionrate}
\Sigma = Q_1\ Q_2^{\ 1-\qS}\ D_k^{\ \qS-4.8}\ (S/N)^{\ -\eta\ (1-\qS)}.
\end{equation}

The telescope apertures required to obtain the corresponding S/N
values for apparent magnitudes of V=8, 10, and 12
mag\footnote{Through-put of 0.65 was used to account for quantum
  efficiency (0.85), mirror reflectivity and transparency of optics
  (0.85), and atmospheric transparency (0.9).}  are shown above the
figure for reference.

% degeneracies and rates in the Main belt
\subsection{Size-Distance Degeneracy and the Main Belt Occultation Rate}
\label{ssec:degen}

Similar sized objects (in Fsu) at different distances will produce
diffraction profiles that are scaled in width with respect to one
another.  This makes it possible for distant and nearby occulters to
produce identical occultation lightcurves; a distant occulter will
cast a wider shadow but the observer will move through it more
quickly.  Here, we demonstrate how such a size/distance conspiracy
manifests itself observationally, and the possibility of degeneracy
with the Main Belt Asteroids (MBAs) is considered.

A diffraction-dominated occultation will have a peak-to-peak width of
$(6\lambda d)^{1/2} \approx 3.5$ Fsu \citep{nihei07}.  The duration of
an occultation observed at wavelength, $\lambda$, for an occulter at
distance, $d$, can be expressed as:

\begin{equation}
  \label{eq:occTime}
  t_{occ} = \frac{ (6\lambda d)^{1/2} }{v_\perp},
\end{equation}

\noindent where $v_\perp$ is the perpendicular velocity of the
occulter described by equation~\ref{eq:vperp}, and the distance to the
occulter is given by:

\begin{equation}
  \label{eq:cosineLaw}
  d = \left[ \rearth^2 + \rocc^2 + 2 \rearth \rocc 
    \cos\left(\varepsilon + \sin^{-1} 
      \left( \frac{ \sin(\varepsilon)}{ r_{\mathrm{o}}} \right) 
    \right) \right]^{1/2}.
\end{equation}

\noindent Here, $r_{\oplus,\mathrm{o}}$ are the Earth-Sun and
occulter-Sun distances, and $\varepsilon$ is the solar elongation of
the target star.

Figure~\ref{fig:occTime} shows occultation time as a function of the
distance to the occulter for targets at different solar elongations.
At opposition, a diffraction-dominated occultation by an object at 40 AU
produces a shadow width similar to that observed for an occultation by
an object at 0.08 AU, just outside of Earth's orbit.  The nearby object
would have to have the same size, in Fsu, to project a shadow with the
same structure; and the ratio of sizes is given by the ratio of
Fresnel scales at the two distances: $(0.08/40)^{1/2} = 0.045$.  Thus,
a 500m diameter KBO at 40 AU will produce an occultation shadow
identical to that produced by a 0.045 $\times$ 500 m = 22m diameter
object at a 0.08 AU.  This is the only degenerate point for
$\varepsilon = 180\arcdeg$; all other objects at all other distances
will produce shadows which are different in either duration or
amplitude.

\middfig{\figSDdegen}

The degenerate point moves to greater distances when $\varepsilon \ne
180$.  The greatest risk of degeneracy occurs when the degenerate
point is within the asteroid belt ($\sim2.0 - 3.5$ AU).  This occurs
when a target is observed at elongations $131\arcdeg \lesssim
\varepsilon \lesssim 141\arcdeg$.  For elongations $\varepsilon <
133\arcdeg$, a second set of degenerate points enters the parameter
space. These points represent possible prograde occultations, as the
elongation has moved beyond the stationary point for the distance in
question.  The second set of degeneracy points lies within the Main
Belt for elongations $116\arcdeg \lesssim \varepsilon \lesssim
125\arcdeg$.

At $\varepsilon\approx136\arcdeg$, the degeneracy point is within the
Main Asteroid Belt at distance $d=1.7$ AU (orbital radius 2.5 AU).
There, a 500 m KBO would be degenerate with a $(1.7
\mathrm{AU}/40 \mathrm{AU})^{1/2} \times$ 500 m = 104 m MBA.

To estimate the number of conflicting occulters we compute the ratio
$P_\mathrm{MBA}/P_\mathrm{KBO}$, with $P=2bv\Sigma$
(equation~\ref{eq:2bvs}).  The shadow widths, $2b$, were taken to be
3.5 Fsu (equation~\ref{eq:width}), and velocities were computed with
equation~\ref{eq:vperp}.  KBO sky surface densities were calculated
with equation~\ref{eq:NoverS}, and MBA sky surface densities were
converted from the power law model of the cumulative luminosity
function (CLF) for the Main Belt with slope $\alpha=0.27$
\citeGladman:

\begin{equation}
  \label{eq:clfmba}
  \Sigma(<m_{\mathrm{R}}) = 210 \times 10^{\alpha\ (m_{\mathrm{R}}-23)}, \ \ 20 < m_{\mathrm{R}} < 23.
\end{equation}

When an object with diameter $D$, albedo $p=0.05$, and heliocentric
distance $\Delta$ is observed in reflected light at phase angle
$\beta=\arcsin( d\sin(\varepsilon)/\Delta )$ (Sine Law) at a distance
$d$ from the Earth, its apparent V magnitude is \citep{cox00}:

\begin{eqnarray}
  V &=& H(\beta) + 5 \log (d \Delta),\\
  H(\beta) &=& H - 2.5 \log[ (1-G)\Phi_1(\beta) + G\Phi_2(\beta)],\\
  \Phi_i &=& \exp[-A_i [\tan(\beta/2)]^{B_i}],\\
  H &=& 15.645 - 2.5\log p - 5 \log D.
\end{eqnarray}

\noindent The constant terms have values: $A_1$ = 3.33, $A_2$=1.87,
$B_1$=0.63, and $B_2$=1.22; and a slope parameter of $G=0.2$ (typical
for MBAs) was used.  A value of V--R=0.4 (typical for MBAs) was then
used to obtain apparent R magnitudes to apply in
equation~\ref{eq:clfmba}.

The size distributions of both the KBO and MBA populations are modeled
as power laws, and the probability ratio,
$P_\mathrm{MBA}/P_\mathrm{KBO}$, remains constant as a function of
size, $D$, when the slopes are equal: $\qS = 5\alpha + 1$ = 2.35.  In
this case the probability ratio is $P_\mathrm{MBA}/P_\mathrm{KBO} =
2.5$, and occultations by MBAs are expected to be more numerous.  If
$\qS < 2.35$ for the Kuiper Belt, occultation events will be
predominantly by Main Belt occulters.

Via this transformation, a 100 m MBA has magnitude
$m_{\mathrm{R}}$=26.6 mag, and is an extrapolation of $\sim$3.6
magnitudes beyond the observed limit of the CLF \citeGladman
The assessment remains valid for MBAs with sizes $D_{\mathrm{MBA}}
\gtrsim$ 500 m, and $D_{\mathrm{KBO}} \gtrsim$ 2500 m.

\subsubsection{Breaking the Size-Distance Degeneracy}
\label{sssec:breakSD}

As the slopes for the size distributions of small KBOs and MBAs are
not well understood, it is worthwhile to examine a method of breaking
the size-distance degeneracy.  

In an occultation survey, a target star with large projected diameter
is undesirable as it causes deterioration of the diffraction profile.
This deterioration could allow a shadow cast by a distant occulting
object to be distinguished from its near-field degenerate counterpart.
Deterioration of a profile due to the size of the background star
becomes significant when the star's radius is comparable to the
Fresnel scale.  The Fresnel scale increases as the square-root of the
distance, but the projected size of the background star increases
linearly with distance to the occulter.  Thus, a star which has a
`large' ($\sim$1 Fsu) projected size at 40 AU has a relatively small
($\sim$0.3 Fsu) projected size at 1.7 AU.  The ideal target star to
break the degeneracy will be the brightest available with a projected
diameter of $\sim1$ Fsu.

% False Positives Section
%

\subsection{False Positives}
\label{ssec:falsepos}

The diffraction-dominated occultation shadows being sought in a
photometric time series can be similar in structure to the noise, and
random statistical fluctuations could be misidentified as
occultations.  Here we evaluate the false positive rate with our
cross-correlation detection algorithm, and compare it to the expected
occultation rate to determine an appropriate detection threshold.

Artificial time series were generated with different noise properties
(S/N, and power-spectral slope), and the cross-correlation detection
algorithm was run on the simulated data.  Three hundred detection
kernels for a point-source with $50\mm < \rKBO < 1200 \mm$, $10 \mAU <
\dKBO < 160 \mAU$, and $b < 3$ Fsu (BKW) were used.
Each time series had length $2^{23} \approx 8.4\times10^6$ points
(58.2 hours at 40 $\invs$ sampling) and was guaranteed not to contain
any real occultation events.  The distribution of false positive
events in standard deviation units is presented in
Figure~\ref{fig:sigmaDistrib} with the expected Gaussian distribution
for reference.

\middfig{\figSigNorm}

The cross-correlation detections are $\sim$10-20$\times$ more common
at all levels of significance because multiple detection kernels were
used.  But, the false positive rate did not increase by a factor of
$300$ (the number of kernels) because the kernels do not represent
independent tests.  By design, the spacing of kernel parameters \rKBO,
\dKBO, and $b$ was chosen to allow an event to be detected by multiple
kernels.

The false positive rate decreases as the slope of the
power spectrum becomes steeper.  An occultation lightcurve has a peak
in power at $\sim0.5$ Fsu (see Figure~\ref{fig:fsupower}), and a
negative slope in the power spectrum indicates that more power is
present in the lower frequencies.  The relative amount of power in the
noise band shared by occultation events decreases as the slope of the
power spectrum steepens, and fewer false positives are observed.

Evaluation of the false positive rate must be done on a
per-observation basis with an accurate model of the photometric noise,
and with the same detection algorithm used on the real data.  An
excess of false positives above the Gaussian estimate will be produced
by any method in which multiple passes are made over the data to
extract different patterns.  In the example presented
(Figure~\ref{fig:sigmaDistrib}), reference to an `$n-\sigma$'
cross-correlation peak would be valid for the kernel tested, but would
be a misleading representation of the statistical significance for the
complete collection of kernels.

% how does one select n-sigma for a threshold based on t-total
\subsubsection{Selection of a Detection Threshold}
\label{sssec:thresh}

There is no established detection threshold above which an event
should be considered a viable candidate.  We have used $8\sigma$
throughout this work and will now offer a justification for that
limit.

False positive events should represent a small portion of the number
of observed events, e.g. 5\% for 2$\sigma$ confidence.  To determine a
detection threshold, we consider the ratio of expected genuine events
to false positives events as a function of statistical significance
(in standard deviation units).

The false positive event rate, $\mu_{\mathrm{fp}}$, was computed by
scaling the amplitude of a Gaussian to fit the false positive
distribution shown in Figure~\ref{fig:sigmaDistrib}.  The scaling
compensates for the increased false positive rate produced by the
cross-correlation with multiple kernels.

The expected occultation rate at each level of statistical
significance was estimated with a plant/recover process for a
comprehensive selection of KBOs.  Occultation profiles were tested for
KBOs at 40 AU, observed in visible light, with radii $10\ \mm < \rKBO <
2000\ \mm$ (n=200, $\delta\rKBO=10\ \mm$), and impact parameters $0\ \mm
< b < 8000\ \mm$ (n=100, $\delta b=341\ \mm$).  The statistical
significance of the selected events was measured in artificial 40
$\invs$ time series with S/N=25,100 and $\beta$=1.0 (1/f noise), and the
occultation rate for each grid point was calculated:

\begin{equation}
  \label{eq:Prip}
  \delta \mu(r_i,b_j) = 2\ \delta b\ v\ \left[\Sigma(r_i-\delta r/2) - \Sigma(r_i+\delta r/2)\right].
\end{equation}

Each grid point was binned in statistical significance, and the
rates, $\delta\mu$, were summed in each of the bins:

\begin{equation}
  \label{eq:Psigma}
  \mu(\sigma_k) = \sum_{i,j} \delta P(r_i,b_j), \quad \mathrm{(\sigma-\delta\sigma/2 < \sigma_k < \sigma+\delta\sigma/2)}.
\end{equation}

The occultation rates for false positives,
$\mu_{\mathrm{fp}}(\sigma_k)$, and expected genuine positives,
$\mu(\sigma_k)$, are shown in Figure~\ref{fig:sigselect} (upper panel)
for different slopes of the KBO size distribution, $\qS$=2,3.  The ratio
of the rates, $\mu(\sigma_k):\mu_{\mathrm{fp}}(\sigma_k)$, is shown in
the lower panel.

\middfig{\figSigSelect}

Variability in the expected occultation rate is due almost entirely to
uncertainty in the slope of the size distribution, $\qS$
(factors of $\gtrsim100\times$ versus $\lesssim10\times$ for changes in S/N).

From Figure~\ref{fig:sigselect}, we see that the expected genuine
occultation rate exceeds the false positive rate for statistical
significances of $\gtrsim7\sigma$ ($\qS$=3), or $\gtrsim8\sigma$
($\qS$=2).  A limit of $8\sigma$ has been used throughout this work.

If the photometric noise properties are sufficiently well understood,
the distribution of false positives can be modeled with artificial
time series.  It would then be possible to infer the presence of
genuine events as a statistical excess among more numerous false
positives.

% discussion section
%
\section{Discussion}
\label{sec:discussion}

The parameter space for occultation studies is vast, including
properties of the occulter (size, distance, sky surface density), the
target star (solar elongation, projected diameter), the time series
(sampling rate, S/N, power spectrum slope), and the stochastic nature
of the event itself (impact parameter).  We have endeavored to reduce
the complexity of this parameter space and determine optimum values
for those parameters which are under an observer's control.  Here, we
discuss other considerations associated with SSO work.

% real noise is log-normal for variance > 0.1
%Much of our analysis has made use of artificial photometric time
%series that were generated to reproduce both the standard deviation of
%the variates, and slope of their power spectrum.  The method we
%describe in Section~\ref{ssec:scintillation} assumes that our
%photometric variates, $h_i$, are Gaussian, rather than log-normal.
%This is valid for variance $(\sigma_h/\overline{h})^2 \lesssim 0.1$
%\citep{dravins97} and we have restricted ourselves to evaluating time
%series with variances well below this limit.  Noise levels approaching
%the variance of 0.1 would be uninteresting for occultation work.

% used same kernel to plant as recovery ... contrived?
The plant and recover method was adopted for a variety of tests
performed, and used the same occultation profile for planting as was
used for construction of the detection kernel.  When used for detection
in real data, multiple kernels are used to cover the parameter space
of interest.  The parameters of an occultation event (\rKBO, \dKBO,
and $b$) would not be expected to match any of the kernels exactly,
but would be recovered due to overlap in their ranges of sensitivity.
Use of the same profile for both planting and detection improved the
efficiency of our testing code, but does not compromise the validity
of any results.

% monochrome diff patterns have power > 1 per Fsu
Our finding that occultation events are best sampled at $f_s$=2
$\invFsu$ is based on 95\% of the cumulative power being in
frequencies below $\fny$=1 $\invFsu$.  This is true only for
diffraction profiles integrated over a range of wavelengths.  An
occultation profile observed in monochromatic light will contain
extended ringing and retain considerable power at frequencies beyond 1
$\invFsu$.  The sampling rate we provide here is valid for filters with
$\sim100$ nm pass-bands, typical of broad-band filters used in optical
astronomy.

% used Poisson noise for sample rate numerical test
Estimates of the occultation rate as a function of the sampling rate
were based on plant/recover tests in time series of Poisson variates
rather than $1/f$ noise.  Our photometric time series were sampled at 40
$\invs$ and we did not extrapolate the power spectrum to higher
frequencies in order to simulate the noise at higher sampling rates.
The tests were repeated with $1/f$ time series for frequencies up to
$f$=40 $\invs$, and these were found to be consistent with the Poisson
noise tests for the range of frequencies compared.

% rate way lower than roques et al . ... why
The detectable occultation rates shown in Figure~\ref{fig:exptime} are
considerably lower than earlier estimates.  An occultation by a
$D\approx$1000 m object would be expected at a rate of
$\mu=10^{-8}-10^{-10}\ \invs$, where the previous rate was
$5\times10^{-7}\ \invs$ (R\&M).  The rates presented here take into
account a break in the size distribution which was not well
established when earlier estimates were made.  The break to a
shallower size distribution at $\rKBO\sim25$ km dramatically reduces
the expected sky surface density of km-sized KBOs.

% \todo{all those fps in bick08? ... hmmm explain }\\
The cross-correlation detection algorithm was used in an occultation
search of 40 $\invs$ photometric time series from two B9V stars in M35
(BKW).  In real photometric time series, false positives
were detected at a higher rate than would be predicted in
Section~\ref{ssec:falsepos} of this work.  The critical difference
between observed and simulated time series is that photometric
stability cannot be maintained for an observed time series.
Atmospheric transparency can change on time scales of a few minutes,
and thereby change the signal to noise ratio of the photometry.  To
guard against this, detection algorithms should normalize a candidate
event's statistical significance by a more local measure of the standard
deviation than has been used in the past -- within seconds, rather than
minutes of a candidate point.

%  Summary Section
%

\section{Summary}
\label{sec:summary}

We have provided a comprehensive assessment of the various observing
parameters relevant in an SSO survey.  Our aim has been to
identify optimal values for parameters to guide future observations
in this field.  The results we have obtained are summarized here.

\begin{itemize}

\item
We have estimated occultation event rates based on a broken power law
size distribution, and found rates of $10^{-8}-10^{-10}\ \invs$ for
km-sized objects.  The variability is due to the large uncertainty in
the size distribution slope, $\qS$, for small objects.

\item
Fourier analysis indicates that 95\% of the cumulative power in the
power spectrum of a $\lambda = 400-700$ nm occultation profile 
is present in frequencies of $f \lesssim1\ \invFsu$, indicating that an
event can be critically (Nyquist) sampled at a rate of $f_s$=2
$\invFsu$.  To probe the Kuiper Belt at 40 AU, this corresponds to a
sampling rate of 40 $\invs$.  Numerical tests of the sampling rate
indicate that sampling significantly above the Nyquist limit
($\gtrsim$80-100 $\invs$) increases read-noise, and reduces the
detectable KBO occultation rate.

\item
The photometric noise present in 40 $\invs$ ground-based observations
has a power spectrum with the form $1/f^\beta$, with $\beta\approx1$.
We have used Fourier methods to generate $1/f$ noise artificially and
simulate realistic photometric time series.  Plant-and-recover tests
were performed with these time series to numerically test various
modes of observation.

\item
When optimally sampling, the detectable occultation event rate is
maximized by observing a target at solar opposition (within $\pm
20-30\arcdeg$).  Observing off-opposition does not significantly
improve detectability of smaller, more numerous objects unless the
sampling rate of the camera is sub-critical for an observation
at opposition.

\item
In a diffraction-dominated occultation event which is well sampled,
the occulter distance will manifests itself in the width (duration) of
the event.  A given distance is degenerate with either one, or three
others, depending on the solar elongation of the observation.
Occultation events by KBOs at 40 AU are degenerate with 2.0--3.5 AU
MBAs for observations performed at elongations of $116\arcdeg \lesssim
\varepsilon \lesssim 125\arcdeg$, or $131\arcdeg
\lesssim\varepsilon\lesssim 141\arcdeg$.

\item
Occultations by MBAs may occur at a rate comparable to those for KBOs
depending on the slopes of the size distributions for the two
populations.  An MBA occultation observed at solar opposition will be
$\sim0.5\times$ the duration of a KBO event, and the two should be
distinguishable.  The sampling rate adopted for a survey targeting
multiple distances should be the highest of the critical sampling
rates for each individual distance ($\sim80\ \invs$ for MBAs).

\item 
Due to the expected rarity of KBO occultation events, false positives
are likely to be more common than real events for levels of
statistical significance $\lesssim7-8\sigma$, depending on the slope
of the KBO size distribution.

\end{itemize}

\acknowledgments This research used the facilities of the Canadian
Astronomy Data Centre (CADC) operated by the National Research Council
(NRC) of Canada with the support of the Canadian Space Agency (CSA).
The research was also supported by a Discovery Grant to DLW by the
Natural Sciences and Engineering Research Council of Canada (NSERC).
Special thanks go to the Herzberg Institute of Astrophysics (HIA) for
their work in the development of the high-speed camera which was used
to perform our observations.  We are grateful to the Davies Foundation
and the Fund for Astrophysical Research (FAR) for their financial
support (\url{foundationcenter.org/grantmaker/fundastro/}).  Finally,
our thanks go to Paul Bourke and Gerhardt Pratt for their
comments on 1/f noise, to Robert Lupton for many helpful
conversations, and to the anonymous referee for valuable advice.  {\it
  Facilities:} \facility{DAO}, \facility{HIA}, \facility{CADC}.

\appendix
\section{Derivation of the $1/f$-Noise Normalization Constant}
\label{app:Anorm}

The normalization constant, $A$ (equations~\ref{eq:real},
and~\ref{eq:imag}) is need to produce an artificial time series with
$1/f$ scintillation-like properties.  With variables as defined by
equations~\ref{eq:real} through~\ref{eq:H0} in
Section~\ref{sssec:generating}, the value of $A$ can be determined
as follows:

Taking the time variate $h_k=\overline{h}+\varepsilon_k$
($\overline{\varepsilon}\equiv 0$,
$\sigma_\varepsilon\equiv\sigma_h$), substitution of
equation~\ref{eq:H0} into Parseval's theorem yields:

\begin{eqnarray}
  \label{eq:appAnorm}
  \sum_{k=0}^{N-1} |h_k|^2 &=&
  \frac{1}{N} \sum_{n=0}^{N-1} |H_n|^2,\\ \nonumber
  \sum_{k=0}^{N-1} (\overline{h} + \varepsilon_k)^2 &=&
  \frac{1}{N} \left[H_0^2 + \sum_{n=1}^{N-1} (A f_n^{\beta/2}a_n)^2\right],\\ \nonumber
  \sum_{k=0}^{N-1} \overline{h}^2 + \sum_{k=0}^{N-1} 2\overline{h}\varepsilon_k + \sum_{k=0}^{N-1} \varepsilon_k^2 &=&
  \frac{1}{N} \left[H_0^2 + A^2 \sum_{n=1}^{N-1} f_n^\beta a_n^2\right],\\ \nonumber
  N^2\overline{h}^2 + 2\overline{h} N^2\overline{\varepsilon} + N^2\sigma_h^2 &=&
  H_0^2 + A^2 \sum_{n=1}^{N-1} f_n^\beta a_n^2,\\ \nonumber
  A &=& N \sigma_h \left[ \sum_{n=0}^{N-1} f_n^\beta a_n^2 \right]^{-1/2}.
\end{eqnarray}

\section{The Average Object Diameter in a Broken Power Law Size Distribution}
\label{app:Dbar}

Equation~\ref{eq:meanwidth} describes the average
diameter of an object in a broken power law size distribution.  The
expression can be derived as follows:

\begin{equation}
  \label{eq:Ameanwidth}
  \overline{D} = \frac{1}{N} \int_{D_0}^{\infty} \frac{\dif N(D)}{\dif D} D\ \dif D = 
  \frac{1}{N} \left[ \int_{D_0}^{D_k} \frac{\dif N(D<D_k)}{\dif D} D\ \dif D + 
  \int_{D_k}^{\infty} \frac{\dif N(D>D_k)}{\dif D} D\ \dif D \right],
\end{equation}

\noindent where $N$ is as defined in equation~\ref{eq:surfaceDensity}.
These two integrals can be expressed as:

\begin{eqnarray}
  \label{eq:Aint1}
  \int_{D_0}^{D_k} \frac{\dif N(D<D_k)}{\dif D} D\ \dif D &=&
  \int_{D_0}^{D_k} \left[ \frac{A (r_{\mathrm{max}}^{1-c} - r^{1-c}_{\mathrm{min}})}{(1-c)(\qL-1)} D_k^{\qS-\qL}\ \left( (1-\qS) D^{-\qS}\right) \right] D \ \dif D, \nonumber \\
  &=&\frac{A(r_{\mathrm{max}}^{1-c} - r^{1-c}_{\mathrm{min}})}{(1-c)(\qL-1)} D_k^{\qS-\qL}  \frac{ \left[ (1-\qS) \left(D_k^{2-\qS} - D_o^{2-\qS}\right) \right]}{2-\qS},
\end{eqnarray}

\noindent and

\begin{eqnarray}
  \label{eq:Aint2}
    \int_{D_0}^{D_k} \frac{\dif N(D>D_k)}{\dif D} D\ \dif D &=&
    \int_{D_k}^{\infty} \left[ \frac{A (r_{\mathrm{max}}^{1-c} - r^{1-c}_{\mathrm{min}})}{(1-c)(\qL-1)}\ \left( (1-\qL) D^{-\qL}\right) \right] D_k \ \dif D, \nonumber\\
    &=& \frac{A (r_{\mathrm{max}}^{1-c} - r^{1-c}_{\mathrm{min}})}{(1-c)}\ \frac{D^{2-\qL}}{\qL-2}.
\end{eqnarray}

\noindent With substituting of equations~\ref{eq:Aint1}, and
\ref{eq:Aint2}, equation~\ref{eq:Ameanwidth} can be reduced to give:

\begin{equation}
  \label{eq:Ameanwidthfinal}
  \overline{D}(\qS,\qL\ne 2) = \left[ \left( \frac{1-\qL}{2-\qL} - \frac{1-\qS}{2-\qS} \right) \left( \frac{D_0}{D_k}\right)^{\qS-2} + \frac{1-\qS}{2-\qS} \right] D_0, \qquad D_0 < D_k.
\end{equation}

\noindent Equation~\ref{eq:Ameanwidthfinal} is undefined for $\qL=2$,
but this is well below current estimates of its value
\citep[$\qL\approx4.8$][]{fuentes08,fraser09}.  The equation, as shown, is
also undefined for $\qS=2$, but has a finite limiting value.  It can
then be rearranged to isolate the $(\qS - 2)$ terms, and take the
limit as $\qS \rightarrow 2$:

\begin{equation}
  \label{eq:limit}
  \overline{D}(\qS=2) = \lim_{\qS\to 2}
  \left[ 
    \left( \frac{1-\qL}{2-\qL} \left(\frac{D_0}{D_k}\right)^{\qS-2} + 
      (1-\qS) \left( \frac{1-(D_0/D_k)^{\qS-2}}{2-\qS} \right)
      \right) 
    \right] D_0.
\end{equation}

The limit can then be computed with L'Hopital's rule by taking derivatives
of the numerator, $(1-(D_0/D_k)^{\qS-2})$, and denominator, $(2-\qS)$, in
the second term of Equation~\ref{eq:limit}:

\begin{eqnarray}
  \label{eq:lhopital}
  \overline{D}(\qS=2) &=& 
  \left[ 
    \left( \frac{1-\qL}{2-\qL} + 
      \lim_{\qS\to 2} (1-\qS) \left( \frac{(D_0/D_k)^{\qS-2} \ln(D_0/D_k)}{-1} \right)
      \right) 
    \right] D_0,\\
    &=&
    \left[ \frac{1-\qL}{2-\qL} + \ln(D_k/D_0) \right] D_0, \qquad D_0 < D_k.
\end{eqnarray}

\bibliographystyle{apj}
\bibliography{bickerton}

\tailfig{\figFTnoise}
\tailfig{\figPowerSpec}
\tailfig{\figFTfake}

\tailfig{\figOccrates}
\tailfig{\figFsupower}

\tailfig{\figOccrateVsElong}
\tailfig{\figDminVsElong}

\tailfig{\figSampRate}
\tailfig{\figNoiseLevel}
\tailfig{\figSDdegen}
\tailfig{\figSigNorm} 
\tailfig{\figSigSelect} 

%%%%%%%%%%%%%%%%%%%%%%%%%%%%%%%%%%%%%%%%%%%%%% table 1

\end{document}